\documentclass{PoS}
\usepackage{graphicx,amsmath,alltt,axodraw2}

\DeclareSymbolFont{ttoperators}{OT1}{cmtt}{m}{n}
\newcommand\Code[1]{{%
  \mathcode`\"="0\the\symttoperators22%
  \mathchardef\$="4\the\symttoperators24%
  \mathcode`\(="4\the\symttoperators28%
  \mathcode`\)="5\the\symttoperators29%
  \mathcode`\-="4\the\symttoperators2D%
  \mathcode`\/="0\the\symttoperators2F%
  \mathcode`\[="4\the\symttoperators5B%
  \mathcode`\]="5\the\symttoperators5D%
  \mathchardef\{="4\the\symttoperators7B%
  \mathchardef\}="5\the\symttoperators7D%
  \ensuremath{\mathtt{#1}}}}

\newcommand\Var[1]{\ensuremath{\mathit{#1}}}
\newcommand\eg{e.g.\ }
\newcommand\ie{i.e.\ }

\title{New Developments in FormCalc 8.4}
\ShortTitle{FormCalc 8.4}

\author{Christoph Gro\ss \\
Institut f\"ur Theoretische Physik und Astrophysik, Universit\"at 
W\"urzburg, D--97074 W\"urzburg, Germany \\
E-mail: chrisgr90@gmx.net}

\author{\speaker{T.~Hahn} \\
MPI f\"ur Physik, F\"ohringer Ring 6, D--80805 Munich, Germany \\
E-mail: hahn@mpp.mpg.de}

\author{Sven Heinemeyer \\
Instituto de F\'isica de Cantabria (CSIC-UC), E-39005 Santander, Spain \\
E-mail: Sven.Heinemeyer@cern.ch}

\author{Federico von der Pahlen \\
Instituto de F\'isica, Universidad de Antioquia, Calle 70 No. 52-21, 
Medell\'in, Colombia \\
E-mail: fp@gfif.udea.edu.co}

\author{Heidi Rzehak \\
Albert-Ludwigs-Universit\"at Freiburg, Physikalisches Institut,
D--79104 Freiburg, Germany \\
E-mail: Heidi.Rzehak@cern.ch}

\author{Christian Schappacher \\
Institut für Theoretische Physik, Karlsruhe Institute of Technology,
D--76128 Karlsruhe, Germany (former address) \\
E-mail: schappacher@kabelbw.de}

\abstract{We present new developments in FeynArts 3.9 and FormCalc 8.4,
in particular the MSSMCT model file including the complete one-loop 
renormalization, vectorization/parallelization issues, and the interface 
to the Ninja library for tensor reduction.
\\ \hbox to\hsize{\hfill Report MPP-2014-263, FR-PHENO-2014-009}
}

\FullConference{Loops and Legs in Quantum Field Theory -- 12th DESY 
Workshop on Elementary Particle Physics,\\
April 27--May 2, 2014 \\
Weimar, Germany}

\begin{document}

\section{Introduction}

FeynArts \cite{FeynArts} and FormCalc \cite{FormCalc} are Mathematica 
packages for the generation and simplification of Feynman diagrams.  In 
a first step, FormCalc reads diagrams up to one-loop order generated by 
FeynArts and returns a simplified symbolic result which can be inspected 
or modified using regular Mathematica commands as well as several 
FormCalc functions and options for analytic manipulations, such as the 
application of Fierz identities.  For efficient numerical evaluation, 
\eg of the cross-section, the analytical expressions can then be written 
out as highly optimized subroutines in Fortran or C, complemented by a 
library of utility functions.  The generated code is largely 
self-contained.  It uses simple Fortran programs to set the inputs 
rather than (static) parameter cards, which would be limited to a 
certain class of models.  The numeric calculation can also be interfaced 
back to Mathematica through MathLink, such that \eg the cross-section 
becomes available as a Mathematica function of its input parameters.

This note presents new features and improvements in FeynArts 3.9 and 
FormCalc 8.4:
\begin{itemize}
\item The MSSMCT model file including the complete one-loop 
renormalization of the complex MSSM.

\item Run-time selection of renormalization schemes.

\item Vectorization/parallelization issues in FormCalc, LoopTools, and 
Cuba.

\item Ninja interface for FormCalc.
\end{itemize}

\section{The MSSMCT Model File}

The full renormalization prescriptions of the MSSMCT model file 
\cite{MSSMCT} cannot be reproduced in this limited space.  Rather, we focus 
on a few conceptually important issues which also required changes in 
FeynArts and FormCalc.

\subsection{The Higgs-boson Sector}

Higher-order corrections are phenomenologically very important in the 
Higgs sector.  Indiscriminately including them on propagators and 
vertices risks upsetting the relations necessary for the proper 
cancellation of UV and IR divergences, however, since this mixes orders 
in perturbation theory.  The masses on the Higgs propagtors should 
ideally be consistent with the Higgs mixing angle $\alpha$ 
parameterizing the vertices, but this conflicts with the fact that 
mixing can occur between all three states $h$, $H$, $A$ at loop level, 
which is not expressible through a single angle $\alpha$.

We opted therefore to formulate the vertices with the tree-level 
$\alpha$ and insert tree-level masses on loop propagators but use 
loop-corrected masses on all other propagators.  At the level of the 
Feynman rules it is not possible to generally avoid incomplete 
cancellations due to a mismatch between tree-level and loop-corrected 
masses, though there are typically process-specific solutions.  Our 
recommendation is to test UV and IR finiteness with loop-corrected 
masses and revert to tree-level masses as far as necessary.

On the technical side, FeynArts needed to be extended to allow a 
detailed choice of the masses inserted on different types of 
propagators.  For example, the declaration of the light CP-even Higgs 
boson $h$ includes two mass specifiers
\begin{verbatim}
  S[1] == {
    Mass -> Mh0,
    Mass[Loop] -> Mh0tree, ... }
\end{verbatim}

\subsection{The Fermion/Sfermion Sector}

The model file is presently limited to minimal flavor violation in the 
sfermion sector, which means that for non-trivial CKM matrix there is a 
slight imbalance between fermions and sfermions; for example, the 
$b$-quark has an admixture from $d$ and $s$ while the $\tilde b$ does 
not.  Because this violates delicate supersymmetric relations, reactions
involving squarks (in particular external ones) may not become finite 
and we have therefore chosen to turn CKM mixing off by default.  It can 
be enabled by setting \Code{\$CKM = True}.

\subsection{Run-time Renormalization Scheme Selection}

In the sfermion and chargino/neutralino sectors the available degrees of 
freedom do not allow all physical quantities to be chosen on-shell 
simultaneously.  Solving for the dependent parameters typically yields 
results which become numerically unstable in certain regions of 
parameter space, \eg because of some combination of mixing-matrix 
elements in the denominator.  For example, in the CCN schemes (two 
charginos and one neutralino on-shell) the renormalization constants 
$\delta M_2$ and $\delta\mu$ diverge in regions where $|\mu|\approx
|M_2|$ (gaugino--Higgsino mixing close to maximal).

For this reason one would like to switch between different schemes at 
run-time, say while scanning over parameter space.  Since the dependent 
parameters vary from scheme to scheme, straightforward if-statements 
such as \Code{dMUE1 = IndexIf[\Var{cond},\,\delta\mu^A,\,\delta\mu^B]} 
may contain overlapping dependencies, however, and cannot naively be 
ordered for evaluation in Fortran.  For example,
\begin{alltt}
dMUE1 = IndexIf[{\it{}cond}, \(\delta\mu\sp{A}(\delta{}M\sb1\sp{A})\), \(\delta\mu\sp{B}\)\,];
dMino11 = IndexIf[{\it{}cond}, \(\delta{}M\sb1\sp{A}\), \(\delta{}M\sb1\sp{B}(\delta\mu\sp{B})\)\,];
\end{alltt}
cannot be ordered as it stands, for $\delta\mu$ is a function of $\delta 
M_1$ in scheme $A$, but oppositely in $B$.
The solution was to have FormCalc join the `if' and `else' parts of 
all if-statements with identical conditions, upon which the statements 
in each branch can be ordered correctly:
\begin{alltt}
IndexIf[ {\it{}cond},
  dMino11 = \(\delta{}M\sb1\sp{A}\);
  dMUE1   = \(\delta\mu\sp{A}(\delta{}M\sb1\sp{A})\),
(* else *)
  dMUE1   = \(\delta\mu\sp{B}\);
  dMino11 = \(\delta{}M\sb1\sp{B}(\delta\mu\sp{B})\) ]
\end{alltt}
On the user side, scheme switching is available in MSSMCT as \eg
\begin{alltt}
$InoScheme = IndexIf[{\it{}cond}, CNN[2,1,3], CCN[1]]
\end{alltt}
where \textit{cond} might be \Code{Abs[Abs[MUE] - Abs[Mino2]] < 50}.

Note that a renormalization-scheme switch in principle requires a 
corresponding transition of the affected parameters from one scheme to 
the other for a fully consistent interpretation of the results.

In the light of recent activities to automate the generation of Feynman 
rules also for one-loop counter terms \cite{NLOCT}, it should clearly be 
stated that the MSSMCT model file goes far beyond simple 
$\overline{\text{DR}}$ renormalization for the BSM part.  As outlined 
above, SUSY relations and restrictions due to the available degrees of 
freedom are taken into account, and renormalization conditions can be 
chosen to obtain physically meaningful renormalized parameters over the 
entire parameter space.  We believe this much wider program requires 
physical understanding and cannot easily be automated.


\section{Vectorization/Parallelization Issues}

\subsection{Vectorization of the Helicity Loop}

The assembly of the squared matrix element in FormCalc can be sketched 
as in the following figure, where the helicity loop sits at the center 
of the calculation:
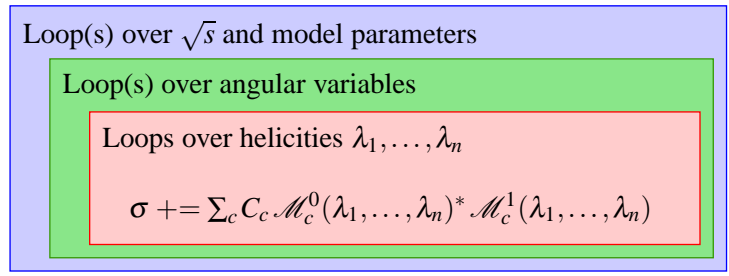
\begin{center}
\begin{picture}(270,100)(0,20)
\CBox(0,20)(270,120){Blue}{PastelBlue}
\Text(5,115)[tl]{Loop(s) over $\sqrt s$ and model parameters}
\CBox(15,25)(265,100){OliveGreen}{PastelGreen}
\Text(20,95)[tl]{Loop(s) over angular variables}
\CBox(30,30)(260,80){Red}{PastelRed}
\Text(35,75)[tl]{Loops over helicities $\lambda_1,\dots,\lambda_n$}
\Text(45,50)[tl]{$\sigma \mathrel{{+}{=}}
  \sum_c C_c\,\mathcal{M}^0_c(\lambda_1,\dots,\lambda_n)^*
            \,\mathcal{M}^1_c(\lambda_1,\dots,\lambda_n)$}
\end{picture}
\end{center}
The helicity loop is not only strategically the most desirable but also 
the most obvious candidate for concurrent execution, as FormCalc does 
not insert explicit helicity states during the algebraic simplification 
\cite{FCopt}.  Such a design is known as Single Instruction Multiple 
Data (SIMD) in computer science and is conceptually easy to vectorize. 
With an AVX-capable i7, a speedup of 3.7 out of a theoretical 4 for the 
helicity loop could be achieved.

Vectorization is available in both Fortran (using Fortran 90's vector 
syntax) and C99 (through macros emitting explicit SSE3/AVX instructions) 
and is enabled by default according to the capabilities of the machine 
on which \Code{configure} is run.  To configure the code for an 
arbitrary machine (\ie without vectorization), the \Code{--generic} flag 
must be used with \Code{configure}.  More in detail, the automatically 
determined maximum hardware vector length is written to \Code{simd.h} 
and the user can control in \Code{distrib.h} whether to include 
\Code{simd.h} (default) or define a different vector length in the 
\Code{SIMD} preprocessor variable (zero for no vectorization).

Since the OPP libraries are currently not capable of handling 
vector-valued numerator functions, \Code{SIMD} must be set to zero in 
\Code{distrib.h} when using the CutTools \cite{CutTools}, Samurai 
\cite{Samurai}, or Ninja \cite{Ninja} libraries.

\subsection{Dropping Negligible Helicity Combinations}

The helicity sum can further be improved by observing that typically 
many terms contribute to the final result only negligibly.  While 
helicity combinations that give an exact zero result could in principle 
be identified analytically, this is not really straightforward due to 
the way the amplitudes are composed of abbreviations, and it is far less 
straightforward for small but non-zero (\eg mass-suppressed) 
combinations.

The FormCalc-generated code adopts a numerical strategy instead: It 
observes the terms in the helicity sum for a (smallish) number of 
phase-space points, then identifies and suppresses computation of the 
negligible ones for all further phase-space points.

Two environment variables control the procedure: \Code{FCHSELN} 
specifies the number $n$ of phase-space points to be observed before 
taking the decision and \Code{FCHSELEPS} specifies how much smaller the 
contribution of an individual helicity contribution must be relative to 
the maximum to be neglected, \ie a helicity combination $\{\lambda\}$ is 
dropped if
$$
X_{\{\lambda\}} < \varepsilon
  \max\limits_{\{\lambda\}} X_{\{\lambda\}}
\quad\text{where}\quad
X_{\{\lambda\}} = \sum\limits_{i=1}^n
  |\mathcal{M}^{(i)}_{\{\lambda\}}|\,.
$$

\subsection{Concurrency Issues in LoopTools}

LoopTools \cite{FormCalc} uses a cache to speed up calculation.  
Originally this was done for tensor coefficients only, for which there 
is both a significant overlap of intermediate results in the computation 
and a high chance that once one coefficient is called, related ones will 
be needed as well.  Recently the cache system was extended to scalar 
integrals, too, so that \Code{C0(\dots)} (formerly not cached) and 
\Code{C0i(cc0,\,\dots)} (formerly cached) are now identical.  Depending 
on the application, the cache can easily make an order-of-magnitude 
difference in run time.

On the other hand, LoopTools was not thread-safe because of the cache. 
(Indeed, the actual computation of the scalar integrals in Fortran with 
its use of common blocks etc.\ was not really thread-safe either.)  In a 
concurrent environment this is most dangerous, as it may \emph{silently} 
lead to wrong results.

Thread-safety has been achieved by serializing cache writes through 
mutexes.  It was also for this reason, to make them thread-safe, that 
the scalar functions were moved into the cache system.  The uncached 
versions (not thread-safe) are still available as \Code{Aputnocache}, 
\Code{Bputnocache}, \Code{C0nocache}, \Code{D0nocache}, 
\Code{E0nocache}.

With serialization, cache management has become a strategically even 
more important task.  Although LoopTools performs efficient binary-tree 
cache lookups, indiscriminate build-up of the cache does not only cost 
memory but performance as well.  Care should be taken therefore to flush 
the cache whenever a lookup will not reasonably be successful anymore, 
\eg when moving to a different point in phase or parameter space, using 
the \Code{clearcache}, \Code{markcache}, and \Code{restorecache} calls.

\subsection{Vectorization and GPU Issues in Cuba}

Cuba \cite{Cuba} is a library for multidimensional numerical 
integration.  From version 3 on it automatically parallelizes sampling 
of the integrand using \Code{fork}/\Code{wait}.  The Cuba calls were 
extended in version 3.3 to include a new argument, \Code{nvec}, through 
which one can have the integrator deliver more than one point per call 
to the integrand and \eg vectorize the computation this way.

Another interesting option is to do massively parallel sampling on a 
GPU.  Re-writing the integrand function for execution on the GPU is 
necessarily a task of the user (\ie cannot be automatically done by 
Cuba).  GPUs have a somewhat different thread model than the CPU, 
however, and so the strategy of determining the slices to be sampled by 
each core in Cuba 3.x was barely sufficient (at least not very 
convenient) for use with a GPU.  Cuba 4.0 redresses this by 
distinguishing `accelerators' (GPU cores) and `cores' (CPU cores, same 
as in 3.x) and distributes samples differently for either kind; most 
importantly it divides a core's share of points into batches no larger 
than the GPU can handle.

Smaller batches are useful even for CPU cores, for they effectively 
load-level the sampling.  To sample 50000 points on four cores, for 
example, Cuba 3.x would distribute 12500 points to each core.  This is 
optimal only if the evaluation of the integrand is equally fast at all 
points, however, otherwise it is smarter to distribute smaller batches 
of a user-defined size, say 2000, which reduces the chance of a 
bottleneck at a very reasonable increase in communication overhead.

\section{Ninja Interface}

Ninja \cite{Ninja} is a library for tensor reduction based on unitarity 
methods.  Unlike the `traditional' OPP libraries CutTools 
\cite{CutTools} and Samurai \cite{Samurai}, Ninja does not rely on 
sampling the numerator alone but performs a Laurent expansion of the 
integral and is thereby able to reconstruct the numerator tensor using 
far fewer samples and in a numerically more stable way.  To pull off 
this feat, Ninja requires several expansions of the numerator function 
$N(q^\mu, \mu^2)$ (with loop momentum $q^\mu$ and renormalization scale 
$\mu$) in the variables $t$, $x$, and $\mu^2$ as follows:
\begin{itemize}
\item the numerator itself: $N(q^\mu, \mu^2)$,

\item the $\mu$ expansion, for computation of the box coefficients:
$N(t v_\perp^\mu, t^2 v_\perp^2)$,

\item the T3 expansion, for computation of the triangle and tadpole 
coefficients:

$N(v_0^\mu + t v_3^\mu + \frac{\beta + \mu^2}{t} v_4^\mu, \mu^2)$,

\item the T2 expansion, for computation of the bubble coefficients:

$N(v_1^\mu + x v_2^\mu + t v_3^\mu +
\frac{\beta_0 + \beta_1 x + \beta_2 x^2 + \mu^2}{t} v_4^\mu, \mu^2)$.
\end{itemize}
FormCalc generates each of these numerator expansions as an independent 
subroutine which computes the coefficients of $t^j x^k \mu^{2l}$ in the 
ranges of $\{j, k, l\}$ needed by Ninja as a function of the remaining 
parameters $v_i$, $\beta_i$.


\section{Summary}

FeynArts 3.9 (\Code{http://feynarts.de}) and FormCalc 8.4 
(\Code{http://feynarts.de/formcalc}) have many new and improved 
features, most notably the MSSMCT model file including the full one-loop 
renormalization of the MSSM, support for vectorization/parallelization, 
and the interface to the Ninja library.

\medskip
\raggedright

\end{document}